\documentclass[11pt,a4paper]{amsart}

\begin{document}

\title[AG type identities from combinations of characters]
{Andrews-Gordon type identities from combinations of Virasoro 
characters}

\author[Feigin]{Boris Feigin}
\address{Landau Institute for Theoretical Physics,
         Chernogolovka, Institusky Prospekt,
         Moscow region, 142432, Russia.} 
\email{feigin@mmcme.ru}

\author[Foda]{Omar Foda}

\address{Department of Mathematics and Statistics,
         University of Melbourne, 
	 Parkville, Victoria 3010, Australia.}
\email{foda@ms.unimelb.edu.au}

\author[Welsh]{Trevor Welsh}

\address{School of Mathematics,
         University of Southampton,
	 Southampton, SO17 1BJ, United Kingdom.}
\email{taw@maths.soton.ac.uk}

\date{}

\begin{abstract}

For $p\in\{3, 4\}$ and all $p' > p$, with $p'$ coprime to $p$, we 
obtain fermionic expressions for the combination 
$\chi^{p, p'}_{1, s} + q^{\Delta} \chi^{p, p'}_{p-1,s}$ of Virasoro 
($W_2$) characters for various values of $s$, and particular choices 
of $\Delta$. Equating these expressions with known product
expressions, we obtain $q$-series identities which are
akin to the Andrews-Gordon identities. 
For $p=3$, these identities were conjectured by Bytsko.
For $p=4$, we obtain identities whose form is a variation on that 
of the $p=3$ cases. These identities appear to be new.

The case $(p,p')=(3,14)$ is particularly interesting because it relates 
not only to $W_2$, but also to $W_3$ characters, and offers $W_3$ 
analogues of the original Andrews-Gordon identities.
Our fermionic expressions for these characters differ from those of
Andrews {\em et al\/} which involve Gaussian polynomials.

\end{abstract}

\maketitle

\newtheorem{Theorem}{Theorem}[section]
\newtheorem{Corollary}[Theorem]{Corollary}
\newtheorem{Proposition}[Theorem]{Proposition}
\newtheorem{Conjecture}[Theorem]{Conjecture}
\newtheorem{Lemma}[Theorem]{Lemma}
\newtheorem{Example}[Theorem]{Example}
\newtheorem{Note}[Theorem]{Note}
\newtheorem{Definition}[Theorem]{Definition}
                                                                               
\renewcommand{\mod}{\textup{mod}\,}
\newcommand{\wt}{\text{wt}\,}

\newcommand{\field}[1]{\mathbb{#1}}
\newcommand{\C}{\field{C}}
\newcommand{\Z}{\field{Z}}

\newcommand{\T}{{\mathcal T}}
\newcommand{\U}{{\mathcal U}}
\newcommand{\tT}{\tilde{\mathcal T}}
\newcommand{\tU}{\widetilde{\mathcal U}}
\newcommand{\Y}{{\mathcal Y}}
\newcommand{\B}{{\mathcal B}}
\newcommand{\D}{{\mathcal D}}
\newcommand{\M}{{\mathcal M}}
\renewcommand{\P}{{\mathcal P}}
\newcommand{\R}{{\mathcal R}}

\newcommand{\Proof}{\medskip\noindent {\it Proof: }}
\newcommand{\Proofof}[1]{\medskip\noindent {\it Proof of #1: }}
\newcommand{\cqfd}{\hfill $\Box$ \medskip}
\newcommand{\run}{{\mathcal X}}
\newcommand{\owt}{{wt}}
\newcommand{\mwt}{\tilde{wt}}
\newcommand{\ochi}{{\chi}}
\newcommand{\mchi}{\tilde{\chi}}
\newcommand{\pchi}{\ddot\chi}
\newcommand{\tkappa}{\tilde{\kappa}}
\newcommand{\bari}{\overline\imath}
\newcommand{\barj}{\overline\jmath}
\newcommand{\barh}{\overline h}
\newcommand{\boldm}{\boldsymbol{m}}
\newcommand{\oboldm}{\overline{\boldm}}
\newcommand{\boldn}{\boldsymbol{n}}
\newcommand{\boldu}{\boldsymbol{u}}
\newcommand{\oboldu}{\overline{\boldu}}
\newcommand{\bolda}{\boldsymbol{a}}
\newcommand{\bolde}{\boldsymbol{e}}
\newcommand{\boldB}{\boldsymbol{B}}
\newcommand{\boldC}{\boldsymbol{C}}
\newcommand{\boldCC}{\overline{\boldsymbol{C}}}
\newcommand{\boldQ}{\boldsymbol{Q}}
\newcommand{\boldQQ}{\overline{\boldsymbol{Q}}}
\newcommand{\boldN}{\boldsymbol{N}}
\newcommand{\boldmu}{\boldsymbol{\mu}}
\newcommand{\boldnu}{\boldsymbol{\nu}}
\newcommand{\boldDelta}{\boldsymbol{\Delta}}
\newcommand{\wombat}{\rule[-6pt]{0pt}{46pt}}
\newcommand{\qbinom}[2]{{\genfrac{[}{]}{0pt}{}{#1}{#2}}_q}
\newcommand{\qbinomq}[3]{{\genfrac{[}{]}{0pt}{}{#1}{#2}}_{#3}}
\newcommand{\qbinoms}[2]{{\textstyle\genfrac{[}{]}{0pt}{}{#1}{#2}}_q}

\hyphenation{And-rews
             Gor-don
             boson-ic
             ferm-ion-ic
             para-ferm-ion-ic
             two-dim-ension-al
             two-dim-ension-al}

\section{Introduction}\label{IntroSec}

\subsection{Identities and Virasoro Characters}\label{AGSec}

The $q$-series identities of And\-rews--Gordon \cite{andrews74,gordon}
take the form:
\begin{equation}\label{AG}
\sum_{N_1 \ge \cdots \ge N_{k-1} \ge 0}
\frac{ q^{ N_{1}^2 + \cdots + N_{k-1}^2 + N_{i} + \cdots + N_{k-1} }}
     {(q)_{N_{1}-N_{2}} \cdots (q)_{N_{k-2}-N_{k-1}} (q)_{N_{k-1}}}
=\prod_{\begin{subarray}{c}
  n=1\\n\not\equiv0, \pm i \, (\mod 2k+1) \end{subarray}}^{\infty}
\frac{1}{1-q^n},
\end{equation}
where $|q|<1$ and, as usual,
$(q)_0=1$ and $(q)_n=\prod_{i=1}^{n}(1-q^i)$ for $n>0$.
Here $k\ge2$ and $1\le i\le k$.
The $k=2$ cases are the famous Rogers-Ramanujan identities
\cite{rogers,rogers-ramanujan}.

In the past twenty years, it was recognised that \eqref{AG} is
an identity for the (normalised) character $\chi_{1, i}^{2, 2k+1}$ of the
minimal model $M(2, 2k+1)_2$ of the Virasoro algebra.
In fact, the left side of \eqref{AG} has a combinatorial
interpretation in terms of particles which are forbidden to overlap.
It is for this reason that the left side of \eqref{AG} is
termed a {\em fermionic} expression.

The Virasoro minimal models $M(p,p')_2$ are labelled by
coprime integers $p$ and $p'$ for which $1<p<p'$.%
\footnote{$M(p,p')_2$ is often denoted $\mathcal M(p,p')$ or
$\mathcal M^{p,p'}$.}
They contain irreducible modules labelled by $r$ and $s$
with $1\le r<p$ and $1\le s<p'$.
In \cite{feigin-fuchs,rocha-caridi}, the characters of these
modules were calculated to be
$\hat\chi^{p, p'}_{r, s}=
q^{\Delta^{p, p'}_{r, s}} \chi^{p, p'}_{r, s}$,
where the normalised character $\chi^{p, p'}_{r, s}$ is given by:
\begin{equation}\label{bosonic-equation}
\chi^{p, p'}_{r, s}=
{\frac{1}{(q)_\infty}}
\sum_{\lambda=-\infty}^\infty
(
q^{\lambda^2pp'+\lambda(p'r-ps)}
-
q^{(\lambda p+r)(\lambda p'+s)}
),
\end{equation}
with $(q)_{\infty}=\prod_{i=1}^\infty (1-q^{i})$,
and the {\em conformal dimension} $\Delta^{p, p'}_{r, s}$ is given by:
\begin{equation}\label{conformal-equation}
\Delta^{p, p'}_{r, s}=
\frac{(p'r-ps)^2-(p'-p)^2}{4pp'}.
\end{equation}
This expression \eqref{bosonic-equation} for $\chi^{p, p'}_{r, s}$
is of a different nature to those on either side of \eqref{AG}.
It is known as a {\em bosonic} expression.
For later purposes, it will be useful to note that
$\chi^{p,p'}_{r,s}=\chi^{p,p'}_{p-r,p'-s}$ and that
$\chi^{p,p'}_{r,s}|_{q=0}=1$.

Following the recognition that \eqref{AG} is an identity for
$\chi^{2,2k+1}_{1,i}$, it was natural to seek
fermionic expressions for other Virasoro characters.
Gradually, as described in \cite{bms,welsh},
beginning with the pioneering work of the Stony Brook group \cite{kkmm},
an increasingly wide range of characters were tackled in a series of works,
culminating in \cite{welsh} giving fermionic expressions for all
minimal model Virasoro characters.
%
%
In the simplest cases, such as the subset of the $p=3$
cases of $\chi^{p,p'}_{r,s}$
tackled by \cite{andrews84,foda-quano}, the fermionic
expressions are strikingly similar to that in \eqref{AG},
with merely the coefficients of some of the parameters changed.
However, expressions for other characters \cite{kkmm,berkovich-mccoy},
most notably the unitary characters $\chi^{p,p+1}_{r,s}$,
necessitated the summand to have one or more Gaussian polynomial
factors, where as usual, the Gaussian polynomial $\qbinom{P}{N}$
is defined by:
\begin{equation}
\qbinom{P}{N}=
\left\{
  \begin{array}{cl}
    \displaystyle
    \frac{(q)_P}{(q)_N(q)_{P-N}}& \text{if $0\le N\le P$;}\\[3mm]
    0& \text{otherwise.}
  \end{array} \right.
\end{equation}
Such a multi-sum expression is known as a {\em fundamental fermionic form}.
In \cite{welsh}, the fermionic expression for $\chi^{p,p'}_{r,s}$
is generally a sum over a number of fundamental fermionic forms.

Of course, to obtain an identity which is an analogue of \eqref{AG},
it is necessary to find an expression for the character which is
a product similar in nature to that on the right.
In fact, as explained in \cite{christe}, $\chi^{p,p'}_{r,s}$
is a product of terms of the form $(1-q^n)^{-1}$ if and only
if $p=2r$, $p'=2s$, $p=3r$ or $p'=3s$:
\begin{equation} \label{ProdEq1}
\chi^{2r, p'}_{r, s}=
\hskip-1mm
\prod_{\begin{subarray}{c}
  n=1\\n\not\equiv0,\pm rs\,(\mod rp')\end{subarray}}^{\infty}
\hskip-2mm
\frac{1}{1-q^n},
\qquad
\chi^{p, 2s}_{r, s}=
\hskip-1mm
\prod_{\begin{subarray}{c}
  n=1\\n\not\equiv0,\pm rs\,(\mod sp)\phantom{'}\end{subarray}}^{\infty}
\hskip-2mm
\frac{1}{1 - q^n},
\end{equation}
\begin{equation}
\label{ProdEq3}
\chi^{3r, p'}_{r, s}=
\hskip-1mm
\prod_{\begin{subarray}{c}
  n=1\\n\not\equiv0,\pm rs\,(\mod 2rp')\\n\not\equiv\pm2r(p'-s)\,(\mod 4rp')\\
       \end{subarray}}^{\infty}
\hskip-3mm\frac{1}{1-q^n},
\qquad
\hskip-1mm
\chi^{p, 3s}_{r, s}=
\prod_{\begin{subarray}{c}
  n=1\\n\not\equiv0,\pm rs\,(\mod 2sp)\\n\not\equiv\pm2s(p-r)\,(\mod 4sp)\\
       \end{subarray}}^{\infty}
\hskip-3mm\frac{1}{1-q^n}.
\end{equation}
These expressions are easily derived by applying Jacobi's triple
product identity \cite[eq.\ (II.28)]{gasper-rahman}
in the cases $p=2r$ and $p'=2s$, and
Watson's quintuple product identity \cite[ex.\ 5.6]{gasper-rahman}
in the cases $p=3r$ and $p'=3s$, to the expression \eqref{bosonic-equation}.
On identifying one of these product expressions with a
fermionic expression for the same $\chi^{p,p'}_{r,s}$,
we obtain an identity which we refer to as a $M(p,p')_2$-identity.

\subsection{Combinations of Virasoro characters}
                                                                                
In \cite{bytfri}, the sum and difference of certain pairs of
characters were considered.
Firstly, use of Watson's quintuple product identity
and \eqref{bosonic-equation} leads to the following
expression (\cite[eqn.~(2.26)]{bytfri}):
\begin{equation}\label{Composep3Eq1}
\chi^{3,p'}_{1,s}\pm q^{\frac{p'}4-\frac s2}\chi^{3,p'}_{2,s}=
 \frac{(\mp q^{\frac{p'}4-\frac s2},\mp q^{\frac{p'}4+\frac s2},
                     q^{\frac{p'}2};q^{\frac{p'}2})_\infty
       (q^{s},q^{p'-s};q^{p'})_\infty}
         {(q)_\infty},
\end{equation}
for $p'\not\equiv0\,(\mod3)$ and $1\le s<p'$,
where
$(a_1,a_2,\ldots,a_t;z)_\infty=
\prod_{j=1}^t (a_j;z)_\infty$
with
$(a;z)_\infty=\prod_{i=0}^\infty (1-az^i)$ as usual.
It will be useful to note the following alternative form of this
expression when $p'\ne2s$:
\begin{equation}\label{Composep3Eq2}
\chi^{3,p'}_{1,s}\pm q^{\frac{p'}4-\frac s2}\chi^{3,p'}_{2,s}=
\frac{
  (q^{s},q^{\frac{p'}2-s},q^{\frac{p'}2};q^{\frac{p'}2})_\infty}
 {(q)_\infty
    (\pm q^{\frac{p'}4-\frac s2},\pm q^{\frac{p'}4+\frac s2};
                                         q^{\frac{p'}2})_\infty}.
\end{equation}

Secondly, use of Jacobi's triple product identity
and \eqref{bosonic-equation} leads to the following
expression (\cite[eqn.~(2.22)]{bytfri}):
\begin{equation}\label{Composep4Eq2}
\chi^{4,p'}_{1,s}\pm q^{\frac{p'}2-s}\chi^{4,p'}_{3,s}=
\frac{(q^{s},\mp q^{\frac{p'}2-s},\mp q^{\frac{p'}2};
                                    \mp q^{\frac{p'}2})_\infty}
         {(q)_\infty},
\end{equation}
for $p'\not\equiv0\,(\mod2)$ and $1\le s<p'$.
This may also be written:
\begin{equation}\label{Composep4Eq1}
\chi^{4,p'}_{1,s}\pm q^{\frac{p'}2-s}\chi^{4,p'}_{3,s}=
\frac{(\mp q^{\frac{p'}2-s},\mp q^{\frac{p'}2},\mp q^{\frac{p'}2+s},
      q^{s},q^{p'-s},q^{p'};q^{p'})_\infty}
         {(q)_\infty}.
\end{equation}

In what follows, we derive fermionic expressions for the \lq$+$\rq\
cases of the above character combinations.
This leads to analogues of the
Andrews-Gordon identities.

In recognition of its origin in a character combination,
we refer to the identity for each sum
$\chi^{3,p'}_{1,s}+ q^{\frac{p'}4-\frac s2}\chi^{3,p'}_{2,s}$
of characters, as an $M(3,p')^+_2$-identity.
Similarly, the identity for each sum
$\chi^{4,p'}_{1,s}+ q^{\frac{p'}2-s}\chi^{4,p'}_{3,s}$
of characters, is referred to as an $M(4,p')^+_2$-identity.

\subsection{\boldmath $M(3,7)_3$-identities}

For $n>2$, the $W_n$ algebra \cite{zamo,fatlyk} is a generalisation
of the Virasoro algebra $W_2$. In \cite{asw}, Andrews, Schilling and 
Warnaar applied a modified Bailey transform to obtain identities
(referred to in \cite{asw} as Rogers-Ramanujan type identities)
for three (of the possible four) characters of the minimal model
$M(3, 7)_3$ of $W_3$.\footnote{One of these identities 
is also proved in \cite{warnaar} using Hall-Littlewood polynomials.}
In Section \ref{ASWSec} below,
we list these identities together with a conjectured identity for
the fourth character.
Notably, the summand on the fermionic side of each of these identities
contains a Gaussian polynomial.

In attempting to understand these $M(3,7)_3$-identities, we noticed that,
since the $W_n$ minimal model $M(p,p')_n$ has central charge
\begin{equation}\label{Ccharge}
c^{p,p'}_n = (n-1)\left(1 - \frac{n(n+1)(p' - p)^2}{p p'}\right),
\end{equation}
the $W_3$ minimal model $M(3, 7)_3$ has the same central 
charge as that of the $W_2$ minimal model $M(3, 14)_2$. 
This indicated that the characters of the $M(3, 7)_3$ theory should
be expressible as linear combinations of those of $M(3, 14)_2$.
In fact, the required combinations are precisely those of
\eqref{Composep3Eq1} for $s\in\{1,3,5\}$,
together with the single character $\chi^{3,14}_{1,7}$.
In each of the first three cases, the known fermionic expressions for
the characters $\chi^{3,14}_{1,s}$ and $\chi^{3,14}_{2,s}$
each comprises a single fundamental fermionic form \cite{bms,welsh},
and sum conveniently to yield a single fundamental fermionic form.
In the latter case, the known fermionic expression
is a sum of two fundamental fermionic forms \cite{welsh}.
However, they sum conveniently to yield a single fundamental fermionic form
in a manner that is similar to the other cases.

\subsection{\boldmath $M(p,p')^+_2$-identities}

Since the above means of combining $W_2$ characters extends readily to
many $p=3$ and $p=4$ cases,
the statement and proof of the $M(3,14)^+_2$-identities
have been subsumed into the general case.
The general $M(3,p')^+_2$-identities and $M(4,p')^+_2$-identities are
stated in Sections \ref{Resp3Sec} and \ref{Resp4Sec} respectively.
Their proofs are given in Section \ref{ProofSec}.

The $M(3,p')^+_2$-identities that we have obtained were conjectured 
by Bytsko \cite{bytsko}. Our $M(4,p')^+_2$-identities are (to the
best of our knowledge) new.

As explained above, the $M(3,14)^+_2$-identities are also identities 
for the $M(3,7)_3$ characters. In order to compare our 
$M(3,14)^+_2$-identities with
the identities of \cite{asw} listed in Section \ref{ASWSec},
we write out the former explicitly in Section \ref{NewSec}.
We see that we now have two different fermionic expressions
for each of the four characters.
In each case, the two fermionic expressions are not simply
related to one another.

\section{Andrews-Gordon type identities}\label{ResSec}

In this section, we list our results.
In each identity, if $s>k$ a sum of the form
$(N_s+N_{s+1}+\cdots+N_k)$ is to be taken as $0$.
Each identity is proved in Section \ref{ProofSec}
using the procedure discussed above.

\subsection{\boldmath $M(3,p')^+_2$-{} and $M(3,p')_2$-identities}
\label{Resp3Sec}

Theorems \ref{FermCase3Cor} and \ref{FermCase1Cor} below give
$M(3,p')^+_2$-identities.
Theorems \ref{FermCase4Cor} and \ref{FermCase2Cor} below give
$M(3,p')_2$-identities.
The $M(3,p')^+_2$-identities originate from the sum of two 
Virasoro characters, one that has $r=1$ and one that has 
$r=2$, as explained above. 
The $M(3,p')_2$-identities originate from one Virasoro character 
only, as can be seen from the proofs in Section \ref{ProofSec}.

In Theorems \ref{FermCase3Cor} and \ref{FermCase1Cor}, the identities 
are indexed by the integers $g$ and $s$. In these two cases, $p'$ is 
related to $g$ by $p'=3g+1$ and $p'=3g+2$ respectively.
In Theorems \ref{FermCase4Cor} and \ref{FermCase2Cor}, the identities 
are indexed by the integer $h$ only. From $h$, one can determine both 
$p'$ and $s$, as explained in Section \ref{ProofSec}. 

\begin{Theorem}\label{FermCase3Cor}
If $g\ge1$ and $1\le s\le g+1$, then:
\begin{equation}
\boxed{
\begin{aligned}
&\sum_{\begin{subarray}{c}
           N_1\ge\cdots\ge N_{g-1}\ge0\\[0.5mm]
           M\ge0
        \end{subarray}}
\hskip-4mm
\frac{q^{\sum_{j=1}^{g-1} N_j(N_j+M)
+\frac {g+1}4 M^2 +\frac {g-s}2 M
+(N_{s}+N_{s+1}+\cdots N_{g-1})}}
{(q)_{N_{1}-N_{2}} \cdots (q)_{N_{g-2}-N_{g-1}} (q)_{N_{g-1}} (q)_M}\\
&\hskip17mm
=\frac{(-q^{\frac{3g+1}4-\frac s2},-q^{\frac{3g+1}4+\frac s2},
                              q^{\frac{3g+1}2};q^{\frac{3g+1}2})_\infty
       (q^s,q^{3g+1-s};q^{3g+1})_\infty}
     {(q)_\infty}.
\end{aligned}
}
\end{equation}
\end{Theorem}

\begin{Theorem}\label{FermCase4Cor}
Let $h\ge1$. Then:
\begin{equation}\label{FermCor4}
\boxed{
\begin{aligned}
&\sum_{\begin{subarray}{c}
           N_1\ge\cdots\ge N_{2h}\ge0\\[0.5mm]
           M\ge0
        \end{subarray}}
\hskip-4mm
\frac{q^{\sum_{j=1}^{2h} N_j(N_j+M)
+\frac{h+1}2 M(M+1)-M
+(N_h+N_{h+1}+\cdots+N_{2h})}}
{(q)_{N_{1}-N_{2}} \cdots (q)_{N_{2h-1}-N_{2h}} (q)_{N_{2h}} (q)_M}\\
&\hskip55mm
=\frac{(q^{3h+2};q^{3h+2})_\infty}{(q)_\infty}.
\end{aligned}
}
\end{equation}
\end{Theorem}

\begin{Theorem}\label{FermCase1Cor}
If $g\ge1$ and $1\le s\le g+1$, then:
\begin{equation}
\boxed{
\begin{aligned}
&\sum_{\begin{subarray}{c}
           N_1\ge\cdots\ge N_{g-1}\ge0\\[0.5mm]
           M\ge0
        \end{subarray}}
\hskip-4mm
\frac{q^{\sum_{j=1}^{g-1} N_j(N_j+M)
+\frac g4 M^2 +\frac {g-s+1}2 M
+(N_{s}+N_{s+1}+\cdots+N_{g-1})}}
{(q)_{N_{1}-N_{2}} \cdots (q)_{N_{g-2}-N_{g-1}} (q)_{N_{g-1}} (q)_M}\\
&\hskip16mm
=\frac{(-q^{\frac{3g}4-\frac{s-1}2},-q^{\frac{3g}4+\frac{s+1}2},
                              q^{\frac{3g}2+1};q^{\frac{3g}2+1})_\infty
       (q^s,q^{3g+2-s};q^{3g+2})_\infty}
     {(q)_\infty}.
\end{aligned}
}
\end{equation}
\end{Theorem}

\begin{Theorem}\label{FermCase2Cor}
Let $h\ge1$. Then:
\begin{equation}
\boxed{
\begin{aligned}
&\sum_{\begin{subarray}{c}
           N_1\ge\cdots\ge N_{2h-1}\ge0\\[0.5mm]
           M\ge0
        \end{subarray}}
\hskip-4mm
\frac{q^{\sum_{j=1}^{2h-1} N_j(N_j+M)
+\frac h2 M(M+1)
+(N_h+N_{h+1}+\cdots+N_{2h-1})}}
{(q)_{N_{1}-N_{2}} \cdots (q)_{N_{2h-2}-N_{2h-1}} (q)_{N_{2h-1}} (q)_M}\\
&\hskip55mm
=\frac{(q^{3h+1};q^{3h+1})_\infty}{(q)_\infty}.
\end{aligned}
}
\end{equation}
\end{Theorem}

We note that, in view of the equality between the right sides
of \eqref{Composep3Eq1} and \eqref{Composep3Eq2},
in the cases in which both $p'$ and $p'/2-s$ are even, the identities
do not involve fractional powers of $q$, with the right side
readily written as a product of terms of the form 
$(1-q^{n})^{-1}$ for various $n\in\Z_{>0}$.
Similarly, in the cases in which $p'$ even and $p'/2-s$ is odd,
the right side is readily written as a product of terms of the form 
$(1-q^{\frac n2})^{-1}$ for various $n\in\Z_{>0}$,
and the identities
may be naturally viewed as identities in the indeterminate $q^{1/2}$,

When $p'$ is odd, the identities may be naturally viewed as identities
in the indeterminate $q^{1/4}$, with in particular, the
product expression able to be readily written as a product of terms of
the forms $(1-q^{\frac n4})^{-1}$ and $(1+q^{\frac m2})^{-1}$
for various $n,m\in\Z_{>0}$.

The identities of Theorems \ref{FermCase4Cor} and \ref{FermCase2Cor},
which are indexed by the integer $h$,
do not involve fractional powers of $q$.
They are $M(3,6h+4)_2$-{} and $M(3,6h+2)_2$-identities respectively.

\subsection{\boldmath $M(4,p')^+_2$-identities}\label{Resp4Sec}

Theorems \ref{FermCase5Cor} and \ref{FermCase6Cor} below
give $M(4,4g+1)^+_2$-identities,
and Theorems \ref{FermCase7Cor} and \ref{FermCase8Cor} below
give $M(4,4g+3)^+_2$-identities.

\begin{Theorem}\label{FermCase5Cor}
Let $g\ge1$ and $1\le s\le g+1$. Then:
\begin{equation}\label{FermCase5CorEq}
\boxed{
\begin{aligned}
&\sum_{\raisebox{0mm}{$
      \begin{subarray}{c}
           N_1\ge\cdots\ge N_{g-1}\ge0\\[0.5mm]
           M\ge0
        \end{subarray}$}}
\frac{q^{\sum_{j=1}^{g-1} N_j(N_j+2m)
+(g+\frac12)M^2+(g-s)M
+(N_s+N_{s+1}+\cdots+N_{g-1})}}
{(q)_{N_{1}-N_{2}} \cdots (q)_{N_{g-2}-N_{g-1}} (q)_{N_{g-1}}
(q^{\frac12};q)_M(q^2;q^2)_M}\\
&\hskip15mm
=\frac{(-q^{2g-s+\frac12},-q^{2g+\frac12},-q^{2g+s+\frac12},
      q^{s},q^{4g+1-s},q^{4g+1};q^{4g+1})_\infty}
         {(q)_\infty}.
\end{aligned}
}
\end{equation}
\end{Theorem}

\begin{Theorem}\label{FermCase6Cor}
Let $g\ge1$. Then:
\begin{equation}\label{FermCase6CorEq}
\boxed{
\begin{aligned}
&\sum_{\raisebox{0mm}{$
      \begin{subarray}{c}
           N_1\ge\cdots\ge N_{g-1}\ge0\\[0.5mm]
           M\ge0
        \end{subarray}$}}
\frac{q^{\sum_{j=1}^{g-1} N_j(N_j+2M+1) + (g+\frac12)M^2+gM}}
{(q)_{N_{1}-N_{2}} \cdots (q)_{N_{g-2}-N_{g-1}} (q)_{N_{g-1}}
(q^{\frac12};q)_{M+1}(q^2;q^2)_M}\\
&\hskip30mm
=\frac{(-q^{\frac12},-q^{2g+\frac12},-q^{4g+\frac12},
      q^{2g},q^{2g+1},q^{4g+1};q^{4g+1})_\infty}
         {(q)_\infty}.
\end{aligned}
}
\end{equation}
\end{Theorem}

\begin{Theorem}\label{FermCase7Cor}
Let $g\ge1$ and $1\le s\le g+1$. Then:
\begin{equation}
\boxed{
\begin{aligned}
&\sum_{\raisebox{0mm}{$
      \begin{subarray}{c}
           N_1\ge\cdots\ge N_{g-1}\ge0\\[0.5mm]
           M\ge0
        \end{subarray}$}}
\frac{q^{\sum_{j=1}^{g-1} N_j(N_j+2M)
+M(Mg+g+1-s)
+(N_s+N_{s+1}+\cdots+N_{g-1})}}
{(q)_{N_{1}-N_{2}} \cdots (q)_{N_{g-2}-N_{g-1}} (q)_{N_{g-1}}
(q^{\frac12};q)_{M}(q^2;q^2)_M}\\
&\hskip15mm
=\frac{(-q^{2g-s+\frac32},-q^{2g+\frac32},-q^{2g+s+\frac32},
      q^{s},q^{4g+3-s},q^{4g+3};q^{4g+3})_\infty}
         {(q)_\infty}.
\end{aligned}
}
\end{equation}
\end{Theorem}

\begin{Theorem}\label{FermCase8Cor}
Let $g\ge1$. Then:
\begin{equation}
\boxed{
\begin{aligned}
&\sum_{\raisebox{0mm}{$
      \begin{subarray}{c}
           N_1\ge\cdots\ge N_{g-1}\ge0\\[0.5mm]
           M\ge0
        \end{subarray}$}}
\frac{q^{\sum_{j=1}^{g-1} N_j(N_j+2M+1) + gM(M+1)}}
{(q)_{N_{1}-N_{2}} \cdots (q)_{N_{g-2}-N_{g-1}} (q)_{N_{g-1}}
(q^{\frac12};q)_{M+1}(q^2;q^2)_M}\\
&\hskip25mm
=\frac{(-q^{\frac12},-q^{2g+\frac32},-q^{4g+\frac52},
      q^{2g+1},q^{2g+2},q^{4g+3};q^{4g+3})_\infty}
         {(q)_\infty}.
\end{aligned}
}
\end{equation}
\end{Theorem}

In each case above, the identities may be viewed as identities in the 
indeterminate $q^{1/2}$, with the product expression readily 
written as a product of terms of the forms
$(1-q^{\frac n2})^{-1}$ and $(1+q^{m})^{-1}$
for various $n,m\in\Z_{>0}$.

\subsection{\boldmath $M(3,7)_3$ Andrews-Gordon identities}\label{NewSec}

The $g=4$ cases of Theorem \ref{FermCase1Cor} for 
$s\in\{1,3,5\}$ and the $h=2$ case of Theorem \ref{FermCase2Cor},
now yield identities for the four $M(3,7)_3$ characters.
Written out in full, these are:

\begin{subequations}
\begin{multline}
\sum_{n_1, n_2, n_3, n_4 \geq 0}
\frac{
q^{
(n_1+n_2+n_3)^2 + (n_2+n_3)^2 + {n_3}^2 + {n_4}^2 
+ (n_1 + 2 n_2 + 3 n_3) n_4
+ (n_1 + 2 n_2 + 3 n_3)
+ 2 n_4
}
}
{(q)_{n_1} (q)_{n_2} (q)_{n_3} (q)_{n_4}}\\ 
=
\prod_{n=1}^{\infty}
\frac{1}{(1-q^{7n-2})(1-q^{7n-3})^2(1-q^{7n-4})^2(1-q^{7n-5})},
\end{multline}

\begin{multline}
\sum_{n_1, n_2, n_3, n_4 \geq 0}
\frac{
q^{
(n_1+n_2+n_3)^2 + (n_2+n_3)^2 + n_3^2 + n_4^2
+ (n_1 + 2 n_2 + 3 n_3) n_4 + n_3 + n_4
}}
{
(q)_{n_1} (q)_{n_2} (q)_{n_3} (q)_{n_4}} 
\\
=
\prod_{n=1}^{\infty}
\frac{1}{(1-q^{7n-1})(1-q^{7n-2})^2 (1-q^{7n-5})^2
(1-q^{7n-6})},
\end{multline}

\begin{multline}
\sum_{n_1, n_2, n_3, n_4 \geq 0}
\frac{q^{
(n_1+n_2+n_3)^2 + (n_2+n_3)^2 + n_3^2 + n_4^2
+ (n_1 + 2 n_2 + 3 n_3) n_4
}}{
(q)_{n_1} (q)_{n_2} (q)_{n_3} (q)_{n_4}
}\\
= 
\prod_{n=1}^{\infty}
\frac{1}
{(1-q^{7n-1})^2(1-q^{7n-3})(1-q^{7n-4})(1-q^{7n-6})^2},
\end{multline}

\begin{multline}
\sum_{n_1, n_2, n_3, n_4 \geq 0}
\frac{q^{
(n_1+n_2+n_3)^2 + (n_2+n_3)^2 + n_3^2 + n_4^2
+ (n_1 + 2 n_2 + 3 n_3) n_4 + (n_2 + 2 n_3) + n_4    
}}{
(q)_{n_1} (q)_{n_2} (q)_{n_3} (q)_{n_4}
}\\ 
=
\prod_{n=1}^{\infty}
\frac{1}
{(1-q^{7n-1})(1-q^{7n-2})(1-q^{7n-3})(1-q^{7n-4})(1-q^{7n-5})(1-q^{7n-6})}.
\end{multline}
\end{subequations}

\subsection{\boldmath $M(3,7)_3$ ASW identities}\label{ASWSec}

In \cite{asw}, Andrews, Schilling and Warnaar obtained
$q$-series identities for three characters of
the $M(3, 7)_3$ minimal model.
These identities differ from those we have derived above.
For comparison purposes,
we list these identities here, together with a conjectured identity
\eqref{ConjASW} for the fourth $M(3, 7)_3$ character:
\begin{subequations}
\begin{multline}
\sum_{n_1, n_2 \geq 0}
\frac{q^{n_1^2 - n_1 n_2 + n_2^2 + n_1 + n_2}}{(q)_{n_1}}
\qbinom{2n_1}{n_2} \\
=
\prod_{n=1}^{\infty}
\frac{1}{(1-q^{7n-2})(1-q^{7n-3})^2(1-q^{7n-4})^2(1-q^{7n-5})},
\end{multline}

\begin{multline}\label{ConjASW}
\sum_{n_1, n_2 \geq 0}
\frac{q^{n_1^2 - n_1 n_2 +n_2^2 + n_2}}{(q)_{n_1}}
\qbinom{2 n_1 + 1}{n_2} \\
=
\prod_{n=1}^{\infty}
\frac{1}{(1-q^{7n-1})(1-q^{7n-2})^2 (1-q^{7n-5})^2 (1-q^{7n-6})},
\end{multline}

\begin{multline}
\sum_{n_1, n_2 \geq 0}
\frac{q^{n_1^2 -n_1 n_2 + n_2^2}}{(q)_{n_1}}
\qbinom{2n_1}{n_2} \\
= 
\prod_{n=1}^{\infty}
\frac{1}{(1-q^{7n-1})^2(1-q^{7n-3})(1-q^{7n-4})(1-q^{7n-6})^2},
\end{multline}

\begin{multline}
\sum_{n_1,n_2\geq 0}
\frac{q^{n_1^2-n_1 n_2+n_2^2+n_1}}{(q)_{n_1}}\qbinom{2n_1+1}{n_2}
=
\sum_{n_1,n_2\geq 0}
\frac{q^{n_1^2-n_1 n_2+n_2^2+n_2}}{(q)_{n_1}}\qbinom{2n_1}{n_2}\\
=
\prod_{n=1}^{\infty}\frac{1}{(1-q^{7n-1})(1-q^{7n-2})(1-q^{7n-3})
(1-q^{7n-4})(1-q^{7n-5})(1-q^{7n-6})}.
\end{multline}
\end{subequations}

\subsection{Special cases}\label{Intp3Sec}
Here we note that the simplest cases of the results in Section \ref{ResSec}
yield known identities of Rogers-Ramanujan type.

The $g=1$ cases of Theorem \ref{FermCase3Cor} yield the
$z=q^{\frac12}$ and $z=1$ specialisations of Euler's formula 
\cite[II.2]{gasper-rahman}.

The $g=1$ cases of Theorem \ref{FermCase1Cor} yield the
following identities after substituting $q\rightarrow q^4$:
\begin{subequations}
\begin{equation}\label{RogersEq1}
\sum_{m=0}^\infty
\frac{q^{m(m+2)}}{(q^4;q^4)_m}=
\frac{(-q^3,-q^7,q^{10};q^{10})_\infty (q^4,q^{16};q^{20})_\infty}
     {(q^4;q^4)_\infty},
\end{equation}
\begin{equation}\label{RogersEq2}
\sum_{m=0}^\infty
\frac{q^{m^2}}{(q^4;q^4)_m}=
\frac{(-q,-q^9,q^{10};q^{10})_\infty (q^8,q^{12};q^{20})_\infty}
     {(q^4;q^4)_\infty}.
\end{equation}
\end{subequations}
These identities are equivalent to identities of Rogers \cite{rogers}.
They also appear in \cite[eqns.\ (16) \&\ (20)]{slater}.

The $g=1$ cases of Theorems \ref{FermCase5Cor} and \ref{FermCase6Cor}
yield the following identities after substituting $q\rightarrow q^2$:
\begin{subequations}
\begin{equation}\label{RogersEq3}
\sum_{m=0}^\infty
\frac{q^{3m^2}}{(q;q^2)_m(q^4;q^4)_m}=
\frac{(-q^3,-q^5,-q^7;q^{10})_\infty}
     {(q^4,q^6;q^{10})_\infty},
\end{equation}

\begin{equation}\label{RogersEq4}
\sum_{m=0}^\infty
\frac{q^{3m^2+2m}}{(q;q^2)_{m+1}(q^4;q^4)_m}=
\sum_{m=0}^\infty
\frac{q^{3m^2-2m}}{(q;q^2)_m(q^4;q^4)_m}=
\frac{(-q,-q^5,-q^9;q^{10})_\infty}
     {(q^2,q^8;q^{10})_\infty}.
\end{equation}
\end{subequations}
In an alternative form, \eqref{RogersEq3} and the second identity
in \eqref{RogersEq4} are due to Rogers \cite{rogers}.
They also appear in \cite[eqns.\ (19) \&\ (15)]{slater}.
Bailey \cite{bailey47} also derives \eqref{RogersEq3} and
the two identities \eqref{RogersEq4}.

{}From Theorems \ref{FermCase7Cor} and \ref{FermCase8Cor}, we obtain:
\begin{subequations}
\begin{equation}\label{SelbergEq1}
\sum_{m=0}^\infty
\frac{q^{2m(m+1)}}{(q;q^2)_m(q^4;q^4)_m}=
\frac{(-q^5,-q^7,-q^9;q^{14})_\infty}
     {(q^4,q^6,q^8,q^{10};q^{14})_\infty},
\end{equation}

\begin{equation}\label{SelbergEq2}
\sum_{m=0}^\infty
\frac{q^{2m^2}}{(q;q^2)_m(q^4;q^4)_m}=
\frac{(-q^3,-q^7,-q^{11};q^{14})_\infty}
     {(q^2,q^6,q^8,q^{12};q^{14})_\infty},
\end{equation}

\begin{equation}\label{SelbergEq3}
\sum_{m=0}^\infty
\frac{q^{2m(m+1)}}{(q;q^2)_{m+1}(q^4;q^4)_m}=
\frac{(-q,-q^7,-q^{13};q^{14})_\infty}
     {(q^2,q^4,q^{10},q^{12};q^{14})_\infty}.
\end{equation}
\end{subequations}
These identities are alternative forms of the Rogers-Selberg identities
which are originally due to Rogers \cite{rogers}.
They also appear in \cite[eqns.\ (32), (33) \&\ (31)]{slater}.

\section{Proofs of fermionic combinations}\label{ProofSec}

In this section, we prove the expressions of Section \ref{ResSec}.
In each case, the first step is to extract fermionic expressions for
characters $\chi^{3,p'}_{1,s}$ or $\chi^{4,p'}_{1,s}$ from \cite{welsh}.
Some of these expressions can also be found in \cite{berkovich-mccoy,bms}.
Having obtained the fermionic forms, the required proofs
result upon combining them appropriately and then using
either \eqref{Composep3Eq1} and \eqref{Composep4Eq2}.

In each of the cases below, we use the notation of \cite{welsh}.

\subsection{The $(p,p')=(3,3g+1)$ case}\label{p3SecA}

The continued fraction of $p'/p=(3g+1)/3$ is $[g,3]$.
Then, from \cite[\S1.3]{welsh}, $n=1$, $t=g+1$ and $t_1=g-1$.
The set ${\mathcal T}$ of Takahashi lengths is given by
${\mathcal T}=(1,2,\ldots,g,g+1)$.
The set $\tilde{\mathcal T}$ of truncated Takahashi lengths is
given by $\tilde{\mathcal T}=(1)$.
{}From \cite[\S1.11]{welsh},
$\boldB$ is the $(g-1)\times (g-1)$ matrix
$\boldB=\{B_{j\ell}\}_{1\le j,\ell\le g-1}$
with entries $B_{j\ell}=\min\{j,\ell\}$, and
$\boldCC$ is the $1\times 1$ matrix $(2)$.
The matrix $\boldCC^*$ is the $1\times 1$ matrix $(-1)$.

Throughout this subsection, we take
$N_j=n_j+n_{j+1}+\cdots+n_{g-1}$ for $1\le j\le g-1$,
and the $(g+1)$-dimensional vectors $\bolde_j$ are defined by
$\bolde_j=(\delta_{1j},\delta_{2j},\ldots,\delta_{g+1,j})$
for $0\le j\le g+2$.

\begin{Lemma}\label{FermCase3}
Let $g\ge1$ and $1\le s\le g+1$. Then:
\begin{subequations}
\begin{equation}\label{FermEx3a}
\chi^{3,3g+1}_{1,s}=
\hskip-5mm
\sum_{\begin{subarray}{c}
           n_1,n_2,\ldots,n_{g-1}\in\Z_{\ge0}\\
           m\in\Z_{\ge0}\\
           m\equiv0\,(\mod2)
        \end{subarray}}
\hskip-8mm
\frac{q^{\sum_{j=1}^{g-1} N_j(N_j+m)
+\frac{g+1}4 m^2 +\frac {g-s}2 m
+(N_s+N_{s+1}+\cdots+N_{g-1})}}
{(q)_{n_1}(q)_{n_2}\cdots (q)_{n_{g-1}}(q)_m}
\end{equation}
and
\begin{equation}\label{FermEx3b}
\chi^{3,3g+1}_{2,s}=
q^{-\frac{3g}4+\frac s2-\frac14}
\hskip-5mm
\sum_{\begin{subarray}{c}
           n_1,n_2,\ldots,n_{g-1}\in\Z_{\ge0}\\
           m\in\Z_{\ge0}\\
           m\equiv1\,(\mod2)
        \end{subarray}}
\hskip-8mm
\frac{q^{\sum_{j=1}^{g-1} N_j(N_j+m)
+\frac{g+1}4 m^2 +\frac {g-s}2 m
+(N_s+N_{s+1}+\cdots+N_{g-1})}}
{(q)_{n_1}(q)_{n_2}\cdots (q)_{n_{g-1}}(q)_m}.
\end{equation}
\end{subequations}
Each term $(N_s+N_{s+1}+\cdots+N_{g-1})$ is to be taken as $0$
whenever $s\ge g$.
\end{Lemma}

\Proof
Since with $1\le s\le g+1$, we have $s\in\T$,
and for $r=1$, we have $r\in\tT$, the fermionic form
\cite[(1.16)]{welsh} for $\chi^{p,p'}_{r,s}$ comprises a single
fundamental form \cite[(1.17)]{welsh}.
The summation in \cite[(1.17)]{welsh} is then over integers
$n_1,n_2,\ldots,n_{g-1}$ and $m$
(this last parameter is named $m_{g}$ in \cite[(1.17)]{welsh}),
where the parity of $m$ is restricted.

{}From \cite[\S1.7]{welsh},
we obtain:
\begin{equation*}
\boldsymbol{u}^L=\begin{cases}
\bolde_{g}&\text{if }s=g+1;\\
0&\text{if }s=g;\\
\bolde_{s-1}-\bolde_{g-1}&\text{if }s\le g-1.
\end{cases}
\end{equation*}
{}From \cite[\S1.9]{welsh}, we then get
$\oboldu^L_\flat=(1)$ if $s=g+1$ and
$\oboldu^L_\flat=(0)$ if $s\le g$.
On the other hand, $\boldu^R=\bolde_g$ and
$\oboldu^R_\sharp=(0)$.
For \cite[(1.18)]{welsh}, we then get
$\tilde\boldn=(\tilde n_1,\tilde n_2,\ldots,\tilde n_{g-1})$
where, if $s\in\{g,g+1\}$ then $\tilde n_{g-1}=n_{g-1}+\frac12m$
and $\tilde n_j=n_j$ for $1\le j<g-1$,
and if $1\le s<g$ then $\tilde n_{g-1}=n_{g-1}+\frac12m+\frac12$,
$\tilde n_{s-1}=n_{s-1}-\frac12$
and $\tilde n_j=n_j$ for $1\le j<s-1$ and $s\le j<g-1$.
The expression \cite[(1.17)]{welsh} then yields equation \eqref{FermEx3a}
up to an overall factor, where $m$ is summed over even integers
because here $\oboldu=(0)$ leads to $\boldQQ(\boldu)=(0)$.
The overall factor may be obtained after calculating $\gamma$ as in
\cite[\S1.10]{welsh}.
However, we can bypass this calculation by noting that
the smallest value of the exponent in the numerator of the summand in
equation \eqref{FermEx1a} occurs when $n_1=n_2=\cdots=n_{2g-1}=m=0$.
Since $\chi^{p,p'}_{r,s}|_{q=0}=1$,
it follows that the required factor is simply $1$.

For the $\chi^{3,p'}_{2,s}$ case, we obtain $\boldu^L$ as above.
Using $r=2$ instead of $r=1$ results in
$\boldu^R=\bolde_g+\bolde_{g+1}$.
This yields $\oboldu^R_\sharp=(0)$ as above.
However here, $m\equiv1\,(\mod2)$ in the summation
because $\oboldu=(1)$ leads to $\boldQQ(\boldu)\equiv(-1)$.
Thus, apart from the overall factor, 
the fermionic form \cite[(1.17)]{welsh} for
$\chi^{3,p'}_{2,s}$ differs from the $r=1$ case above only
in that $m$ is summed over odd integers.
The smallest value of the exponent in the numerator of the summand
in equation \eqref{FermEx3b} occurs when
$n_1=n_2=\cdots=n_{2g-1}=0$ and $m=1$.
It follows that the required factor is
$q^{-\frac{3g}4+\frac s2-\frac14}$.
\cqfd

\Proofof{Theorem \ref{FermCase3Cor}}
Combining Lemma \ref{FermCase3} with equation \eqref{Composep3Eq1}
immediately gives the required result.
\cqfd

\begin{Lemma}\label{FermCase4}
Let $h\ge1$ and set $s=3h+2$. Then:
\begin{equation}\label{FermEx4}
\chi^{3,2s}_{1,s}=
\hskip-5mm
\sum_{\begin{subarray}{c}
           n_1,n_2,\ldots,n_{2h}\in\Z_{\ge0}\\
           m\in\Z_{\ge0}
        \end{subarray}}
\hskip-8mm
\frac{q^{\sum_{j=1}^{2h} N_j(N_j+m)
+\frac{h+1}2 m(m+1)-m
+(N_h+N_{h+1}+\cdots+N_{2h})}}
{(q)_{n_1}(q)_{n_2}\cdots (q)_{n_{2h}}(q)_m},
\end{equation}
where $N_j=n_j+n_{j+1}+\cdots+n_{2h}$.
\end{Lemma}

\Proof
Here $p'=6h+4$, and so the above data applies for $g=2h+1$.
Since $s\not\in\T$, the fermionic expression \cite[(1.16)]{welsh} is
a sum over more than one fundamental fermionic form.
In fact, the Takahashi tree for $s$ (\cite[\S1.6]{welsh}) has
precisely two leaf nodes. Since the truncated Takahashi tree for $r=1$
has one leaf node, the sum \cite[(1.16)]{welsh} is over two terms.

For the leaf node $a_{00}$ of the Takahashi tree for $s$,
we obtain (\cite[\S1.7]{welsh}) $d=2$,
$\Delta_1=1$, $\sigma_1=g$, $\tau_1=g+2$,
$\Delta_2=-1$, $\tau_2=g-2$ and $\sigma_2=(g-3)/2$.
This leads to
$\boldu^L=\bolde_{(g-3)/2}-\bolde_{g-2}+\bolde_{g}+\bolde_{g+1}$.
Thereupon, $\oboldu^L_\flat=(1)$.
On the other hand $r=1$ leads to
$\boldu^R=\bolde_{g}$ and $\oboldu^R_\sharp=(0)$.
For this case, the calculation described in (\cite[\S1.10]{welsh})
yields the constant term $\gamma(\chi^L,\chi^R)=-(g-1)/2$.
The fundamental fermionic form \cite[(1.17)]{welsh} for 
this case is then given by precisely the right side of equation 
\eqref{FermEx4}, with the summation restricted to odd integers $m$ 
because $\oboldu=(1)$ here, which yields $\boldQQ(\boldu)\equiv(-1)$.

For the leaf node $a_{10}$ of the Takahashi tree for $s$,
we obtain $d=2$,
$\Delta_1=-1$, $\sigma_1=g-1$, $\tau_1=g+2$,
$\Delta_2=1$, $\tau_2=g-2$ and $\sigma_2=(g-3)/2$.
This leads to $\boldu^L=\bolde_{(g-3)/2}-\bolde_{g-2}+\bolde_{g}$.
Thereupon, $\oboldu^L_\flat=(1)$.
Again we then obtain $\gamma(\chi^L,\chi^R)=-(g-1)/2$.
The fundamental fermionic form for this case
is then given by precisely the right side of equation \eqref{FermEx4},
with the summation restricted to even integers $m$ because
$\oboldu=(0)$ here, which yields $\boldQQ(\boldu)=(0)$.
The expression \eqref{FermEx4} follows.
\cqfd

\Proofof{Theorem \ref{FermCase4Cor}}
Combining Lemma \ref{FermCase4} with the $p'=2s$ case of
equation \eqref{Composep3Eq1}, and cancelling a factor of 2 from
each side, yields the required result.
\cqfd

\subsection{The $(p,p')=(3,3g+2)$ case}\label{p3SecB}

The continued fraction of $p'/p=(3g+2)/3$ is $[g,1,2]$.
Then, from \cite[\S1.3]{welsh}, $n=2$, $t=g+1$, $t_1=g-1$ and $t_2=g$.
The set ${\mathcal T}$ of Takahashi lengths is given by
${\mathcal T}=(1,2,\ldots,g+1)$.
The set $\tilde{\mathcal T}$ of truncated Takahashi lengths is
given by $\tilde{\mathcal T}=(1)$.
{}From \cite[\S1.11]{welsh},
$\boldB$ is the $(g-1)\times(g-1)$ matrix
$\boldB=\{B_{j\ell}\}_{1\le j,\ell\le g-1}$
with entries $B_{j\ell}=\min\{j,\ell\}$, and
$\boldCC$ is the $1\times 1$ matrix $(1)$.
The matrix $\boldCC^*$ is the $1\times 1$ matrix $(-1)$.

Throughout this subsection, we take
$N_j=n_j+n_{j+1}+\cdots+n_{g-1}$ for $1\le j\le g-1$,
and the $(g+1)$-dimensional vectors $\bolde_j$ are defined by
$\bolde_j=(\delta_{1j},\delta_{2j},\ldots,\delta_{g+1,j})$
for $0\le j\le g+2$.

\begin{Lemma}\label{FermCase1}
Let $g\ge1$ and $1\le s\le g+1$. Then:
\begin{subequations}
\begin{equation}\label{FermEx1a}
\chi^{3,3g+2}_{1,s}=
\hskip-5mm
\sum_{\begin{subarray}{c}
           n_1,n_2,\ldots,n_{g-1}\in\Z_{\ge0}\\
           m\in\Z_{\ge0}\\
           m\equiv0\,(\mod2)
        \end{subarray}}
\hskip-8mm
\frac{q^{\sum_{j=1}^{g-1} N_j(N_j+m)
+\frac g4 m^2 +\frac{g-s+1}2 m
+(N_s+N_{s+1}+\cdots+N_{g-1})}}
{(q)_{n_1}(q)_{n_2}\cdots (q)_{n_{g-1}}(q)_m}
\end{equation}
and
\begin{equation}\label{FermEx1b}
\chi^{3,3g+2}_{2,s}=
q^{-\frac{3g}4+\frac s2-\frac12}
\hskip-5mm
\sum_{\begin{subarray}{c}
           n_1,n_2,\ldots,n_{g-1}\in\Z_{\ge0}\\
           m\in\Z_{\ge0}\\
           m\equiv1\,(\mod2)
        \end{subarray}}
\hskip-8mm
\frac{q^{\sum_{j=1}^{g-1} N_j(N_j+m)
+\frac g4 m^2+\frac{g-s+1}2 m
+(N_s+N_{s+1}+\cdots+N_{g-1})}}
{(q)_{n_1}(q)_{n_2}\cdots (q)_{n_{g-1}}(q)_m}.
\end{equation}
\end{subequations}
Each term $(N_s+N_{s+1}+\cdots+N_{g-1})$ is to be taken as $0$
whenever $s\ge g$.
\end{Lemma}

\Proof
Since the expressions here are obtained in a manner very similar to those of
Lemma \ref{FermCase3}, we do not give the details.
We only note that with $1\le s\le g+1$ here,
$s\in\T$ and then:
\begin{equation*}
\boldsymbol{u}^L=\begin{cases}
0&\text{if }s=g+1;\\
-\bolde_{g}&\text{if }s=g;\\
\bolde_{s-1}-\bolde_{g-1}-\bolde_g&\text{if }s\le g-1.
\end{cases}
\end{equation*}
\cqfd

\Proofof{Theorem \ref{FermCase1Cor}}
Combining Lemma \ref{FermCase1} with equation \eqref{Composep3Eq1}
immediately yields the required result.
\cqfd

\begin{Lemma}\label{FermCase2}
Let $h\ge1$ and $s=3h+1$. Then:
\begin{equation}\label{FermEx2}
\chi^{3,2s}_{1,s}=
\hskip-5mm
\sum_{\begin{subarray}{c}
           n_1,n_2,\ldots,n_{2h-1}\in\Z_{\ge0}\\
           m\in\Z_{\ge0}
        \end{subarray}}
\hskip-8mm
\frac{q^{\sum_{j=1}^{2h-1} N_j(N_j+m)
+\frac h2 m(m+1)
+(N_h+N_{h+1}+\cdots+N_{2h-1})}}
{(q)_{n_1}(q)_{n_2}\cdots (q)_{n_{2h-1}}(q)_m},
\end{equation}
where $N_j=n_j+n_{j+1}+\cdots+n_{2h-1}$.
\end{Lemma}

\Proof
Since the expression here is obtained in a manner very similar to that of
Lemma \ref{FermCase4}, we do not give the details.
We only note that, after setting $g=2h$,
the leaf node $a_{00}$ of the Takahashi tree for $s$
yields $d=2$,
$\Delta_1=1$, $\sigma_1=g$, $\tau_1=g+2$,
$\Delta_2=-1$, $\tau_2=g-1$ and $\sigma_2=g/2-1$,
which leads to
$\boldu^L=\bolde_{g/2-1}-\bolde_{g-1}+\bolde_{g+1}$,
$\oboldu^L_\flat=(0)$ and
$\gamma(\chi^L,\chi^R)=-g/2$;
and the leaf node $a_{10}$ yields $d=2$,
$\Delta_1=-1$, $\sigma_1=g+2$, $\tau_1=g+2$,
$\Delta_2=1$, $\tau_2=g-1$ and $\sigma_2=g/2-1$,
which leads to $\boldu^L=\bolde_{g/2-1}-\bolde_{g-1}$,
$\oboldu^L_\flat=(0)$ and
$\gamma(\chi^L,\chi^R)=-g/2$.
\cqfd

\Proofof{Theorem \ref{FermCase2Cor}}
Combining Lemma \ref{FermCase2} with the $p'=2s$ case of
equation \eqref{Composep3Eq1}, and cancelling a factor of 2 from
each side, yields the required result.
\cqfd

\subsection{The $(p,p')=(4,4g+1)$ case}\label{p4SecA}

In this and the following subsection, we concentrate on the
cases where $p=4$.
In these cases, we make use of the following simple consequences
of the $q$-binomial theorem \cite[II.4]{gasper-rahman}:
%
\begin{subequations}
\begin{align}
\label{CauchyEq1}
&\sum_{k=0}^\infty q^{\frac12 k^2}\qbinom Pk
=(-q^{\frac12};q)_P,\\
\label{CauchyEq2}
&\sum_{k=0}^\infty q^{\frac12 k^2-Pk}\qbinom Pk
=q^{-\frac12 P^2} (-q^{\frac12};q)_P.
\end{align}
\end{subequations}


The continued fraction of $p'/p=(4g+1)/4$ is $[g,4]$.
Then, from \cite[\S1.3]{welsh}, $n=1$, $t=g+2$ and $t_1=g-1$.
The set ${\mathcal T}$ of Takahashi lengths is given by
${\mathcal T}=(1,2,\ldots,g,g+1,2g+1)$.
The set $\tilde{\mathcal T}$ of truncated Takahashi lengths is
given by $\tilde{\mathcal T}=(1,2)$.
{}From \cite[\S1.11]{welsh},
$\boldB$ is the $(g-1)\times(g-1)$ matrix
$\boldB=\{B_{j\ell}\}_{1\le j,\ell\le g-1}$
with entries $B_{j\ell}=\min\{j,\ell\}$, and
$\boldCC$ is the $2\times 2$ matrix
$\left(\begin{smallmatrix}2&-1\\-1&2\end{smallmatrix}\right)$.
The matrix $\boldCC^*$ is the $2\times 2$
matrix $\left(\begin{smallmatrix}-1&2\\0&-1\end{smallmatrix}\right)$.

Throughout this subsection, we take
$N_j=n_j+n_{j+1}+\cdots+n_{g-1}$ for $1\le j\le g-1$,
and the $(g+2)$-dimensional vectors $\bolde_j$ are defined by
$\bolde_j=(\delta_{1j},\delta_{2j},\ldots,\delta_{g+2,j})$
for $0\le j\le g+3$.

\begin{Lemma}\label{FermCase5}
Let $g\ge1$ and $1\le s\le g+1$. Then:
\begin{subequations}
\begin{multline}\label{FermEx5a}
\chi^{4,4g+1}_{1,s}=
\sum_{\raisebox{0mm}{$
      \begin{subarray}{c}
           n_1,n_2,\ldots,n_{g-1}\in\Z_{\ge0}\\
           m_1,m_2\in\Z_{\ge0}\\
           m_1,m_2\equiv0\,(\mod2)
        \end{subarray}$}}
\hskip-8mm \\
\frac{q^{\sum_{j=1}^{g-1} N_j(N_j+m_1)
+\frac{g+1}4 m_1^2+\frac12 m_2^2-\frac12m_1m_2+\frac12(g-s)m_1
+(N_s+N_{s+1}+\cdots+N_{g-1})}}
{(q)_{n_1}(q)_{n_2}\cdots (q)_{n_{g-1}}(q)_{m_1}}
\qbinom{\frac12 m_1}{m_2}
\end{multline}
and
\begin{multline}\label{FermEx5b}
\chi^{4,4g+1}_{3,s}= 
q^{s-2g-\frac12} 
\hskip-5mm
\sum_{\raisebox{0mm}{$
      \begin{subarray}{c}
           n_1,n_2,\ldots,n_{g-1}\in\Z_{\ge0}\\
           m_1,m_2\in\Z_{\ge0}\\
           m_1\equiv0\,(\mod2)\\
           m_2\equiv1\,(\mod2)
        \end{subarray}$}}
\hskip-8mm \\
\frac{q^{\sum_{j=1}^{g-1} N_j(N_j+m_1)
+\frac{g+1}4 m_1^2+\frac12 m_2^2-\frac12m_1m_2+\frac12(g-s)m_1
+(N_s+N_{s+1}+\cdots+N_{g-1})}}
{(q)_{n_1}(q)_{n_2}\cdots (q)_{n_{g-1}}(q)_{m_1}}
\qbinom{\frac12 m_1}{m_2}.
\end{multline}
\end{subequations}
Each term $(N_s+N_{s+1}+\cdots+N_{g-1})$ is to be taken as $0$
whenever $s\ge g$.
\end{Lemma}

\Proof
Since with $s\le g+1$ and $r=1$, we have $s\in\T$ and $r\in\tT$,
the fermionic expression
\cite[(1.16)]{welsh} for $\chi^{p,p'}_{r,s}$ comprises a single
fundamental fermionic form \cite[(1.17)]{welsh}.
The summation in \cite[(1.17)]{welsh} is then over integers
$n_1,n_2,\ldots,n_{g-1}$ and $m_1$ and $m_2$
(these latter two parameters are named $m_{g}$ and $m_{g+1}$
in \cite[(1.17)]{welsh}), with the parity of $m_1$ and $m_2$ are restricted.

{}From \cite[\S1.7]{welsh},
we obtain:
\begin{equation*}
\boldsymbol{u}^L=\begin{cases}
\bolde_g&\text{if }s=g+1;\\
0&\text{if }s=g;\\
\bolde_{s-1}-\bolde_{g-1}&\text{if }s<g.
\end{cases}
\end{equation*}
{}From \cite[\S1.9]{welsh}, we then get
$\oboldu^L_\flat=(1,0)$ if $s=g+1$ and
$\oboldu^L_\flat=(0,0)$ if $s\le g$.
On the other hand, $\boldu^R=\bolde_g$ and
$\oboldu^R_\sharp=(0,0)$.
For \cite[(1.18)]{welsh}, we then get
$\tilde\boldn=(\tilde n_1,\tilde n_2,\ldots,\tilde n_{g-1})$
where,
if $s\in\{g,g+1\}$ then $\tilde n_{g-1}=n_{g-1}+\frac12m_1$
and $\tilde n_j=n_j$ for $1\le j<g-1$,
and if $1\le s<g$ then $\tilde n_{g-1}=n_{g-1}+\frac12m_1+\frac12$,
$\tilde n_{s-1}=n_{s-1}-\frac12$,
and $\tilde n_j=n_j$ for $1\le j<s-1$ and $s\le j<g-1$.
The expression \cite[(1.17)]{welsh} then yields equation 
\eqref{FermEx5a} up to an overall factor, where $m_1$ and $m_2$ 
are each summed over even integers
because here $\oboldu=(0,0)$ leads to $\boldQQ(\boldu)=(0,0)$.
The overall factor may be obtained by noting that the smallest
value of the exponent in the numerator of the summand in equation 
\eqref{FermEx5a} occurs when $n_1=n_2=\cdots=n_{g-1}=m_1=m_2=0$.
Since $\chi^{p,p'}_{r,s}|_{q=0}=1$,
it follows that the required factor is simply $1$.

For the second expression, we obtain $\boldu^L$ as above. Using 
$r=3$ instead of $r=1$ results in $\boldu^R=\bolde_{g}+\bolde_{g+2}$.
This yields $\oboldu^R_\sharp=(0,0)$ as above. However here, 
$m_1\equiv0\,(\mod2)$ and $m_2\equiv1\,(\mod2)$ in the summation
because $\oboldu=(0,1)$ leads to $\boldQQ(\boldu)\equiv(-2,-1)$.
Thus, apart from the overall factor, the sum form 
\cite[(1.17)]{welsh} for $\chi^{4,4g+1}_{3,s}$ differs from the 
$r=1$ case above only in that $m_2$ is summed over odd integers.
The smallest value of the exponent in the numerator of the summand
in equation \eqref{FermEx5b} occurs when
$n_1=n_2=\cdots=n_{2g-1}=0$ and $m_1=2$ and $m_2=1$.
It follows that the required factor is $q^{s-2g-\frac12}$.
\cqfd

\Proofof{Theorem \ref{FermCase5Cor}}
The expressions \eqref{FermEx5a} and \eqref{FermEx5b} of Lemma \ref{FermCase5}
may be combined to yield a sum over all $m_2\in\Z_{\ge0}$.
On performing this summation using \eqref{CauchyEq2} with $P=\frac12m_1$,
we obtain:
\begin{equation*}
\begin{aligned}
&\chi^{4,4g+1}_{1,s} + q^{2g+\frac12-s} \chi^{4,4g+1}_{3,s}\\
&\hskip5mm
=
\sum_{\raisebox{0mm}{$
      \begin{subarray}{c}
           n_1,n_2,\ldots,n_{g-1}\in\Z_{\ge0}\\
           m_1\in\Z_{\ge0}\\
           m_1\equiv0\,(\mod2)
        \end{subarray}$}}
\hskip-7mm
\frac{q^{\sum_{j=1}^{g-1} N_j(N_j+m_1)
+\frac{2g+1}8 m_1^2+\frac12(g-s)m_1
+(N_s+\cdots+N_{g-1})} (-q^{\frac12};q)_{\frac12m_1} }
{(q)_{n_1}(q)_{n_2}\cdots (q)_{n_{g-1}}(q)_{m_1}}.
\end{aligned}
\end{equation*}
On setting $m_1=2M$ and noting that
$(q)_{2M}=(q;q^2)_M(q^2;q^2)_M
=(-q^{\frac12};q)_M$ $(q^{\frac12};q)_M(q^2;q^2)_M$,
we obtain the left side of \eqref{FermCase5CorEq}.
The right side follows from \eqref{Composep4Eq1}.
\cqfd

\begin{Lemma}\label{FermCase6}
Let $g\ge1$. Then:
\begin{subequations}
\begin{equation}\label{FermEx6a}
\chi^{4,4g+1}_{1,2g+1}=
q^{-\frac{g+1}4}
\hskip-10mm
\sum_{\raisebox{-6.5mm}{$
      \begin{subarray}{c}
           n_1,n_2,\ldots,n_{g-1}\in\Z_{\ge0}\\
           m_1,m_2\in\Z_{\ge0}\\
           m_1\equiv1\,(\mod2)\\
           m_2\equiv0\,(\mod2)
        \end{subarray}$}}
\hskip-10mm
\frac{q^{\sum_{j=1}^{g-1} N_j(N_j+m_1)
+\frac{g+1}4 m_1^2+\frac12 m_2^2-\frac12m_1m_2-\frac12m_2}}
{(q)_{n_1}(q)_{n_2}\cdots (q)_{n_{g-1}}(q)_{m_1}}
\qbinom{\frac12(m_1+1)}{m_2}
\end{equation}
and
\begin{equation}\label{FermEx6b}
\chi^{4,4g+1}_{3,2g+1}=
q^{-\frac{g-1}4}
\hskip-10mm
\sum_{\raisebox{-5mm}{$
      \begin{subarray}{c}
           n_1,n_2,\ldots,n_{g-1}\in\Z_{\ge0}\\
           m_1,m_2\in\Z_{\ge0}\\
           m_1,m_2\equiv1\,(\mod2)
        \end{subarray}$}}
\hskip-10mm
\frac{q^{\sum_{j=1}^{g-1} N_j(N_j+m_1)
+\frac{g+1}4 m_1^2+\frac12 m_2^2-\frac12m_1m_2-\frac12m_2}}
{(q)_{n_1}(q)_{n_2}\cdots (q)_{n_{g-1}}(q)_{m_1}}
\qbinom{\frac12(m_1+1)}{m_2}.\!\!\!\!
\end{equation}
\end{subequations}
\end{Lemma}

\Proof
Here $s=2g+1$ so that $s\in\mathcal T$.
Thereupon, $\boldu^L=\bolde_{g+1}$, ${\oboldu}^L_\flat=(0,1)$
and $\tilde\boldn=(n_1,n_2,\ldots,n_{g-2},n_{g-1}+\frac12m_1)$.
The proof now follows the lines of Lemma \ref{FermCase5},
noting that for $r=1$, we obtain $\boldu=(1,0)$ whereupon
${\boldQQ}(\boldu)=(1,0)$;
and for $r=3$, we obtain $\boldu=(1,1)$ whereupon
${\boldQQ}(\boldu)=(1,1)$.
\cqfd

\Proofof{Theorem \ref{FermCase6Cor}}
After noting that $\chi^{4,4g+1}_{1,2g+1}=\chi^{4,4g+1}_{3,2g}$
and $\chi^{4,4g+1}_{3,2g+1}=\chi^{4,4g+1}_{1,2g}$,
expressions \eqref{FermEx6a} and \eqref{FermEx6b} of Lemma \ref{FermCase6}
may be combined to yield a sum over all $m_2\in\Z_{\ge0}$.
On performing this summation using \eqref{CauchyEq2} with $P=\frac12(m_1+1)$,
we obtain:

\begin{equation*}
\begin{aligned}
&\chi^{4,4g+1}_{1,2g} + q^{\frac12} \chi^{4,4g+1}_{3,2g}\\
&\hskip10mm
=
q^{-\frac{2g-1}8}
\hskip-5mm
\sum_{\raisebox{-5mm}{$
      \begin{subarray}{c}
           n_1,n_2,\ldots,n_{g-1}\in\Z_{\ge0}\\
           m_1\in\Z_{\ge0}\\
           m_1\equiv1\,(\mod2)
        \end{subarray}$}}
\hskip-5mm
\frac{q^{\sum_{j=1}^{g-1} N_j(N_j+m_1)
+\frac{2g+1}8 m_1^2-\frac14m_1} (-q^{\frac12};q)_{\frac12(m_1+1)} }
{(q)_{n_1}(q)_{n_2}\cdots (q)_{n_{g-1}}(q)_{m_1}}.
\end{aligned}
\end{equation*}
On setting $m_1=2M+1$ and noting
$(q)_{2M+1}=(q;q^2)_{M+1}(q^2;q^2)_M
=(-q^{\frac12};q)_{M+1}$ $(q^{\frac12};q)_{M+1}(q^2;q^2)_M$,
we obtain the left side of \eqref{FermCase6CorEq}.
The right side follows from \eqref{Composep4Eq1}.
\cqfd


\subsection{The $(p,p')=(4,4g+3)$ case}\label{p4SecB}

The continued fraction of $p'/p=(4g+3)/4$ is $[g,1,3]$.
Then, from \cite[\S1.3]{welsh}, $n=1$, $t=g+2$, $t_1=g-1$ and $t_2=g$.
The set ${\mathcal T}$ of Takahashi lengths is given by
${\mathcal T}=(1,2,\ldots,g,g+1,2g+1)$.
The set $\tilde{\mathcal T}$ of truncated Takahashi lengths is
given by $\tilde{\mathcal T}=(1,2)$.
{}From \cite[\S1.11]{welsh},
$\boldB$ is the $(g-1)\times(g-1)$ matrix
$\boldB=\{B_{j\ell}\}_{1\le j,\ell\le g-1}$
with entries $B_{j\ell}=\min\{j,\ell\}$, and
$\boldCC$ is the $2\times 2$ matrix
$\left(\begin{smallmatrix}1&1\\-1&2\end{smallmatrix}\right)$.
The matrix $\boldCC^*$ is the $2\times 2$
matrix $\left(\begin{smallmatrix}-1&2\\0&-1\end{smallmatrix}\right)$.

Throughout this subsection, we take
$N_j=n_j+n_{j+1}+\cdots+n_{g-1}$ for $1\le j\le g-1$,
and the $(g+2)$-dimensional vectors $\bolde_j$ are defined by
$\bolde_j=(\delta_{1j},\delta_{2j},\ldots,\delta_{g+2,j})$
for $0\le j\le g+3$.

\begin{Lemma}\label{FermCase7}
Let $g\ge1$ and $1\le s\le g+1$. Then:
\begin{subequations}
\begin{multline}\label{FermEx7a}
\chi^{4,4g+3}_{1,s}=
\hskip-5mm
\sum_{\raisebox{-1mm}{$
      \begin{subarray}{c}
           n_1,n_2,\ldots,n_{g-1}\in\Z_{\ge0}\\
           m_1,m_2\in\Z_{\ge0}\\
           m_1,m_2\equiv0\,(\mod2)
        \end{subarray}$}}
\hskip-8mm\\
\frac{q^{\sum_{j=1}^{g-1} N_j(N_j+m_1)
+\frac g4 m_1^2+\frac12 m_2^2+\frac12(g+1-s)m_1
+(N_s+N_{s+1}+\cdots+N_{g-1})}}
{(q)_{n_1}(q)_{n_2}\cdots (q)_{n_{g-1}}(q)_{m_1}}
\qbinom{\frac12 m_1}{m_2}
\end{multline}
and
\begin{multline}\label{FermEx7b}
\chi^{4,4g+3}_{3,s}=
q^{s-2g-\frac32}
\hskip-5mm
\sum_{\raisebox{-1mm}{$
      \begin{subarray}{c}
           n_1,n_2,\ldots,n_{g-1}\in\Z_{\ge0}\\
           m_1,m_2\in\Z_{\ge0}\\
           m_1\equiv0\,(\mod2)\\
           m_2\equiv1\,(\mod2)
        \end{subarray}$}}
\hskip-8mm\\
\frac{q^{\sum_{j=1}^{g-1} N_j(N_j+m_1)
+\frac g4 m_1^2+\frac12 m_2^2+\frac12(g+1-s)m_1
+(N_s+N_{s+1}+\cdots+N_{g-1})}}
{(q)_{n_1}(q)_{n_2}\cdots (q)_{n_{g-1}}(q)_{m_1}}
\qbinom{\frac12 m_1}{m_2}.
\end{multline}
\end{subequations}
Each term $(N_s+N_{s+1}+\cdots+N_{g-1})$ is to be taken as $0$
whenever $s\ge g$.
\end{Lemma}

\Proof
Since with $s\le g+1$ and $r=1$, we have $s\in\T$ and $r\in\tT$,
the fermionic expression
\cite[(1.16)]{welsh} for $\chi^{p,p'}_{r,s}$ comprises a single
fundamental fermionic form \cite[(1.17)]{welsh}.
The summation in \cite[(1.17)]{welsh} is then over integers
$n_1,n_2,\ldots,n_{g-1}$ and $m_1$ and $m_2$
(these latter two parameters are named $m_{g}$ and $m_{g+1}$
in \cite[(1.17)]{welsh}), with the parity of $m_1$ and $m_2$ are restricted.

{}From \cite[\S1.7]{welsh},
we obtain:
\begin{equation*}
\boldsymbol{u}^L=\begin{cases}
0&\text{if }s=g+1;\\
\bolde_g&\text{if }s=g;\\
\bolde_{s-1}-\bolde_{g-1}-\bolde_g&\text{if }s<g.
\end{cases}
\end{equation*}
{}From \cite[\S1.9]{welsh}, we then get
$\oboldu^L_\flat=(0,0)$ if $s=g+1$ and
$\oboldu^L_\flat=(-1,0)$ if $s\le g$.
On the other hand, $\boldu^R=0$ and
$\oboldu^R_\sharp=(0,0)$.
For \cite[(1.18)]{welsh}, we then get
$\tilde\boldn=(\tilde n_1,\tilde n_2,\ldots,\tilde n_{g-1})$
where,
if $s\in\{g,g+1\}$ then $\tilde n_{g-1}=n_{g-1}+\frac12m_1$
and $\tilde n_j=n_j$ for $1\le j<g-1$,
and if $1\le s<g$ then $\tilde n_{g-1}=n_{g-1}+\frac12m_1+\frac12$,
$\tilde n_{s-1}=n_{s-1}-\frac12$,
and $\tilde n_j=n_j$ for $1\le j<s-1$ and $s\le j<g-1$.
The expression \cite[(1.17)]{welsh} then yields equation 
\eqref{FermEx7a} up to an overall factor, where $m_1$ and $m_2$ 
are each summed over even integers because here $\oboldu=(0,0)$ 
leads to $\boldQQ(\boldu)=(0,0)$.
The overall factor may be obtained by noting that the smallest
value of the exponent in the numerator of the summand in equation 
\eqref{FermEx7a} occurs when $n_1=n_2=\cdots=n_{g-1}=m_1=m_2=0$.
Since $\chi^{p,p'}_{r,s}|_{q=0}=1$,
it follows that the required factor is simply $1$.

For the second expression, we obtain $\boldu^L$ as above.
Using $r=3$ instead of $r=1$ results in
$\boldu^R=\bolde_{g+2}$.
This yields $\oboldu^R_\sharp=(0,0)$, as above.
However here, $m_1\equiv0\,(\mod2)$ and $m_2\equiv1\,(\mod2)$ in the summation
because $\oboldu=(0,1)$ leads to $\boldQQ(\boldu)\equiv(-2,-1)$.
Thus, apart from the overall factor, 
the fermionic form 
for $\chi^{4,4g+1}_{3,s}$ differs from the $r=1$ case above only
in that $m_2$ is summed over odd integers.
The smallest value of the exponent in the numerator of the summand
in equation \eqref{FermEx7b} occurs when
$n_1=n_2=\cdots=n_{2g-1}=0$ and $m_1=2$ and $m_2=1$.
It follows that the required factor is $q^{s-2g-\frac32}$.
\cqfd

\Proofof{Theorem \ref{FermCase7Cor}}
The required result is obtained from Lemma \ref{FermCase7}
and \eqref{CauchyEq2} along the lines used to prove
Theorem \ref{FermCase5Cor}.
\cqfd

\begin{Lemma}\label{FermCase8}
Let $g\ge1$. Then:
\begin{subequations}
\begin{equation}\label{FermEx8a}
\chi^{4,4g+3}_{1,2g+1}=
q^{-\frac{g}4}
\hskip-10mm
\sum_{\raisebox{-6.5mm}{$
      \begin{subarray}{c}
           n_1,n_2,\ldots,n_{g-1}\in\Z_{\ge0}\\
           m_1,m_2\in\Z_{\ge0}\\
           m_1\equiv1\,(\mod2)\\
           m_2\equiv0\,(\mod2)
        \end{subarray}$}}
\hskip-3mm
\frac{q^{\sum_{j=1}^{g-1} N_j(N_j+m_1)
+\frac g4 m_1^2+\frac12 m_2^2}}
{(q)_{n_1}(q)_{n_2}\cdots (q)_{n_{g-1}}(q)_{m_1}}
\qbinom{\frac12(m_1+1)}{m_2}
\end{equation}
and
\begin{equation}\label{FermEx8b}
\chi^{4,4g+3}_{3,2g+1}=
q^{-\frac{g+2}4}
\hskip-10mm
\sum_{\raisebox{-5mm}{$
      \begin{subarray}{c}
           n_1,n_2,\ldots,n_{g-1}\in\Z_{\ge0}\\
           m_1,m_2\in\Z_{\ge0}\\
           m_1,m_2\equiv1\,(\mod2)
        \end{subarray}$}}
\hskip-3mm
\frac{q^{\sum_{j=1}^{g-1} N_j(N_j+m_1)
+\frac g4 m_1^2+\frac12 m_2^2}}
{(q)_{n_1}(q)_{n_2}\cdots (q)_{n_{g-1}}(q)_{m_1}}
\qbinom{\frac12(m_1+1)}{m_2}.
\end{equation}
\end{subequations}
\end{Lemma}

\Proof
Here $s=2g+1$ so that $s\in\mathcal T$.
Thereupon, $\boldu^L=\bolde_{g+1}$, ${\oboldu}^L_\flat=(0,0)$
and $\tilde\boldn=(n_1,n_2,\ldots,n_{g-2},n_{g-1}+\frac12m_1)$.
The proof now follows the lines of the proof of Lemma \ref{FermCase7},
noting that for $r=1$, we obtain $\boldu=(1,0)$ whereupon
${\boldQQ}(\boldu)=(1,0)$;
and for $r=3$, we obtain $\boldu=(1,1)$ whereupon
${\boldQQ}(\boldu)=(1,1)$.
\cqfd

\Proofof{Theorem \ref{FermCase8Cor}}
The required result is obtained from Lemma \ref{FermCase8}
and \eqref{CauchyEq2} along the lines used to prove
Theorem \ref{FermCase6Cor}.
\cqfd

\section{Discussion}

In this paper, for $p=3$ and $p=4$, we have proved fermionic-product
identities for sums $\chi^{p, p'}_{1, s} + q^{\Delta} \chi^{p, p'}_{p-1,s}$
of characters for particular choices of $\Delta$.

For $p>5$, Bytsko and Fring \cite{bytfri} showed that the corresponding 
sums of characters do not have a product form.
Furthermore, a detailed investigation of the fermionic expressions
of \cite{welsh} 
(which is not reported in this work) shows that it is also not possible 
to directly sum such expressions to obtain a single fermionic form.
Thus these $p>5$ cases do not lead to analogues of the Andrews-Gordon
identities.

In \cite{bytfri}, product forms were also derived for 
the difference of two characters when $p=3$, $p=4$ and $p=6$.
In the $p=3$ cases in which $p'$ and $p'/2-s$ are both even, these 
lead to new identities similar to our $M(p, p')^+_2$-identities with 
a sign factor $(-1)^M$ incorporated into the summands.
In the other $p=3$ cases and all of the $p=4$ cases, the identities
so obtained are equivalent to our $M(p, p')^+_2$-identities.

\section*{Acknowledgements}

This work started when two of us (BF and OF) were visiting Kyoto 
University, and Tokyo University, Japan. We thank our colleagues 
there for their hospitality, and Ole Warnaar for discussions. 

BF is partially supported by grant number RFBR 05-01-01007, and
OF by the Australian Research Council (ARC).

\end{document}